\documentclass[aoas,MSNbibl,nameyear,seceqn,dvips]{arximspdf}
\usepackage{graphicx}

\doi{10.1214/12-AOAS538} %kopijuoti is PTS
\volume{6}
\issue{3}
\pubyear{2012}
\firstpage{1047}
\lastpage{1067}

\makeatletter
\def\logit{\operatorname{logit}}
\newtheorem{thmn}{Theorem}[section]
\newcommand{\eqref}[1]{(\ref{#1})}
\makeatother

\begin{document}
\begin{frontmatter}

\title{Detecting mutations in mixed sample sequencing data using empirical Bayes}
\runtitle{Mutation detection by empirical Bayes}

\begin{aug}
\author[a]{\fnms{Omkar} \snm{Muralidharan}\corref{}\thanksref{t1}\ead[label=e1]{omkar@stanford.edu}}, %
\author[b]{\fnms{Georges} \snm{Natsoulis}\thanksref{t2}}, %\ead[label=e2]{}
\author[b]{\fnms{John} \snm{Bell}\thanksref{t3}}, %\ead[label=e4]{}
\author[c]{\fnms{Hanlee} \snm{Ji}\thanksref{t4}} %\ead[label=e4]{}
\and
\author[a]{\fnms{Nancy R.} \snm{Zhang}\thanksref{t5}\ead[label=e5]{nzhang@stanford.edu}}
\thankstext{t1}{Supported in part by an NSF VIGRE Fellowship.}
\thankstext{t2}{Supported in part by the National Institutes of Health
Grants RC2HG005570, R21CA140089.}
\thankstext{t3}{Supported in part by the National Institutes of Health
Grants P01HG000205
RC2HG005570, R21CA140089, U01CS151920.
Also supported by the Howard Hughes Medical Foundation Early Career Grant.}
\thankstext{t4}{Supported in part by the National Institutes of Health
Grants P01HG000205, RC2HG005570, R21CA140089, U01CS151920.
Also supported by the Howard Hughes Medical Foundation Early Career
Grant and The Doris Duke Charitable Foundation.}
\thankstext{t5}{Supported in part by NIH R01 HG006137-01 and NSF DMS
Grant ID 1043204.}
\runauthor{O. Muralidharan et al.}
\affiliation{Stanford University}
\address[a]{O. Muralidharan\\
N. R. Zhang\\
Department of Statistics\\
Stanford University\\
Sequoia Hall\\
390 Serra Mall\\
Stanford, California 94305-4065\\
USA\\
\printead{e5}}

\address[b]{G. Natsoulis\\
J. Bell\\
Stanford Genome Technology Center\\
Stanford University\\
Stanford, California, 94305\\
USA}

\address[c]{H. Ji\\
Division of Oncology\\
Department of Medicine\\
Stanford University School of Medicine\\
Stanford, California 94305-4065\\
USA}

\end{aug}

% HISTORY:
\received{\smonth{8} \syear{2011}}
\revised{\smonth{1} \syear{2012}}

% ABSTRACT
%
\begin{abstract}
We develop statistically based methods to detect single nucleotide DNA
mutations in next generation sequencing data. Sequencing generates
counts of the number of times each base was observed at hundreds of
thousands to billions of genome positions in each sample. Using these
counts to detect mutations is challenging because mutations may have
very low prevalence and sequencing error rates vary dramatically by
genome position. The discreteness of sequencing data also creates a
difficult multiple testing problem: current false discovery rate
methods are designed for continuous data, and work poorly, if at all,
on discrete data.

We show that a simple randomization technique lets us use continuous
false discovery rate methods on discrete data. Our approach is a useful
way to estimate false discovery rates for any collection of discrete
test statistics, and is hence not limited to sequencing data. We then
use an empirical Bayes model to capture different sources of variation
in sequencing error rates. The resulting method outperforms existing
detection approaches on example data sets.
\end{abstract}

% KEYWORDS


\vspace*{-5pt}
\end{frontmatter}

%s1 ###
\section{Introduction}\label{sec1}

Highly-multiplex sequencing technologies have made DNA sequencing
orders of magnitude faster and cheaper [\citet{Shendure2008}].
One\vadjust{\goodbreak}
promising application of next generation sequencing technologies is
detecting changes in the DNA of genetically mixed samples. Examples of this
detection problem include searching for somatic mutations in tumor
tissue contaminated by normal stroma, finding single nucleotide variants
by pooled sequencing of multiple samples, and detecting low-prevalence
mutations in evolving virus populations. Our goal is to find genome positions
at which a fraction of the cells or viruses in the sample have mutated.
The studies we consider are exploratory in nature, so any mutations
we detect will be tested further using more laborious methods.

\citet{Shendure2008} describe the typical sequencing experiment. DNA
from the sample is extracted and fragmented. The fragments are used to
form a DNA library, possibly after amplification and size selection.
The ends of the fragments in the DNA library are sequenced to obtain
fixed-length DNA segments called \textit{reads}. Aligning the reads to
a reference genome yields counts of the number of times each base
$(A,C,G,T)$ is observed at each reference position.
If every cell or virus in the sample has the same base as the reference
genome at a given position, any observed base different from the
reference base must be due to error. Such errors can be caused by
errors in sequencing or alignment.

We define the observed error rate as the proportion of bases observed
at a given position that are not equal to the reference base. For
example, if the reference base at a position were $A$ and we observed
$8$ $A$'s and $2$ $C$'s at the position, the observed error rate would be
$20\%$. Mutations appear in sequencing data as unusually high observed
error rates. For example, suppose we know that the true error rate at a
given genome position is exactly $1\%$. If we observe an error rate of
$2\%$ at that position, and if the total count of all bases observed
for that position is sufficiently high to dismiss sampling noise, then
we can infer that roughly $1\%$ of the cells in the sample carry a
mutation. In practice, we do not know the true error rate, which varies
widely across positions and is affected by many steps in the sequencing
experiment. Also, in most sequencing experiments, a large proportion of
the positions have few counts, making it important to account for
sampling noise. Distinguishing true mutations from uninteresting
randomness requires statistical modeling and analysis.

The discrete nature of sequencing data makes the mixed sample detection
problem particularly challenging. It is difficult to detect small,
continuous changes using discrete data. In addition, sequencing
depth---the total number of $\{ A,C,T,G\} $ counts---varies
dramatically across positions. For example, in targeted resequencing,
the sequencing depth can vary over two to three orders of magnitude
[\citet{Natsoulis2011}, \citet{Porreca2007}]. Any method must work for both low
and high depth positions, which rules out convenient large-sample
approximations.

The discreteness of sequencing data also makes it difficult to tackle
multiple testing issues. False discovery rate $(\mathit{fdr})$
methods are a standard approach to controlling\vadjust{\goodbreak} type I error in
exploratory studies; these methods can be interpreted as empirical
Bayes versions of Bayesian hypothesis tests [\citet
{Benjamini1995}, \citet{Efron2001}, \citet{Efron2004a}]. Current $\mathit{fdr}$ methods, however,
are designed for continuous data, and work poorly on discrete data.

In this paper, we develop an empirical Bayes approach to detect
mutations in mixed samples. First, in Section \ref
{sec:Multiple-Testing-Tools}, we show continuous $\mathit{fdr}$ methods can be
applied to discrete data. Our basic idea is to replace traditional
discrete $p$-values with randomized $p$-values that behave
continuously, and then use continuous $\mathit{fdr}$ methods. It is easy to show
that the resulting method preserves the empirical Bayes interpretation
of false discovery rates. Our approach is a useful way to estimate
false discovery rates for any collection of discrete test statistics,
and is not limited to sequencing data.

Next, in Section~\ref{sec:Modeling-Sequencing-Error}, we present an
empirical Bayes model for sequencing error rates. Mutations appear in
the data as unusually high error rates, so to detect mutations
accurately, we need to estimate the position-wise error distribution
under the null hypothesis of no mutation. We use a hierarchical model
to separate the variation in observed error rates into sampling
variation due to finite depth, variation in error rate at a fixed
position across samples, and variation in error rate across positions.
This model shares information across samples and across genome
positions to estimate the sequencing error rate at each position. We
use the position- and sample-specific null distributions from this
model to screen for mutations.

Finally, in Section~\ref{sec:Results}, we apply our methods to two very
different mutation detection problems. The first problem is motivated
by the detection of emerging mutations in virus samples. We use a
synthetic data set created by \citet{Flaherty2011}, where the truth is
known, to evaluate the accuracy of our method and to make comparisons.
The second problem is the analysis of sequencing data from tumor
samples with matched normal samples. We use this larger and more
complex data set to illustrate the general applicability of our methods.

%s2 ###
\section{Multiple testing tools for discrete data}\label{sec:Multiple-Testing-Tools}

In this section, we show how continuous false discovery methods can be
applied on discrete data. We begin by briefly reviewing the basic steps
in a standard empirical $\mathit{fdr}$ analysis as described by \citet
{Efron2004a}, and showing that none of the steps can be directly
applied to discrete data. We then use a randomization technique to
translate each step to the discrete setting.

%s2.1 ###
\subsection{A continuous false discovery rate analysis}

Consider the following multiple testing problem. We observe continuous
valued data $x_{i},i=1,\ldots,P$, and, based on a model for the null
hypothesis, we have a null distribution~$F_{i}$ for each~$x_{i}$. We
think that most $x_{i}$ are null, and we want to find the few that are
not. For example, our nulls could be normal, $F_{i}=\mathcal{N}
(0,\sigma_{i}^{2})$, and we could be searching for unusually\vadjust{\goodbreak}
large $x_{i}$s. Typically, we use the null distributions to form a
$p$-value for each case:
\[
p_{i}  =  F_{i}(z_{i}).
\]
The $p_i$'s all have the same distribution under the null, since if
$x_{i}\sim F_{i}$, $p_{i}\sim \operatorname{Unif}(0,1)$.

An $\mathit{fdr}$ analysis as outlined by \citet{Efron2004a} proceeds in three
major steps. First, we check the validity of our null distributions. If
our nulls are correct, and most $x_{i}$ are null, then most $x_{i}\sim
F_{i}$. This means that if our nulls are correct, most $p_{i}\sim
\operatorname{Unif}(0,1)$. We can thus use the distribution of the $p_{i}$
to check if our nulls are correct. If they are, the $p$-value histogram
should be uniform through most of the unit interval, possibly with some
extra mass near $0$ and $1$ from truly nonnull~$x_{i}$s. If the
$p$-value histogram has this form, our nulls are at least correct on
average [\citet{Gneiting2007} make this precise].

Often, however, the $p$-value histogram reveals that our null
distributions are wrong. If this happens, our next step is to correct
our null distributions. One way to do this is to estimate the null
using the data [\citet{Efron2004a}]. When our null distributions are
wrong, Efron suggests modeling the null $p$-values as still having a
common distribution, but fitting that distribution using the data
instead of assuming it is $\operatorname{Unif}(0,1)$. Since most of our
hypotheses presumably are null, we can estimate such an ``empirical
null'' by fitting the distribution of the center of the data. We then
use that fitted null distribution to make better $p$-values. If $H:
[0,1]\mapsto[0,1]$ is the cdf of the fitted null
$p$-value distribution, this correction changes our null
distributions~$F_{i}$ to $H\circ F_{i}$ and our $p$-values $p_{i}$ to $H
(p_{i})$.

Finally, once our nulls have been corrected, we can proceed to the
final step of estimating the local false discovery rate
\[
\mathit{fdr}(x_{i})=P(H_{i0}|x_{i}),
\]
where $H_{i0}$ is the event that the $i$th null hypothesis is true.
Using Bayes' rule, and the one-to-one relationship between $x_{i}$ and
the transformed $p$-values, we can express the false discovery rate as
%
%e1 ###
\begin{equation}
\mathit{fdr}(x_{i})=\frac{P(H_{i0}
)f_{\mathrm{null}}(p_{i})}{f(p_{i})},\label{eq:cont-fdr}
\end{equation}
where $f_{\mathrm{null}}(p_{i})$ and $f(p_{i})$ are the
null and marginal distributions of the $p$-values. Note that we can
reasonably model the $p$-values as having the same marginal
distribution because they all have the same distribution under the null.

We estimate the false discovery rate by estimating each of the three
quantities on the right side of (\ref{eq:cont-fdr}). Because we think
that most hypotheses are null, we can simply bound $P(H_{i0})$ by 1,
and since we have corrected our null distributions, we know that
$f_{\mathrm{null}}$ is the uniform density. Last, we can estimate the marginal\vadjust{\goodbreak}
distribution $f$ using the observed $p$-values. Substituting these
quantities into (\ref{eq:cont-fdr}) yields an estimated $\mathit{fdr}$, which we
can use to find nonnull hypotheses based on the magnitudes of the $x_{i}$'s.

%s2.2 ###
\subsection{Discrete data problems}

The three core steps in our continuous false discovery rate analysis
are checking the null distributions, possibly estimating an empirical
null, and estimating $\mathit{fdr}$'s. Each step relies on the assumption that
if we knew the correct null distributions of our test statistics, the
null $p$-values would be uniform. This assumption fails for discrete
data: even when all of our null distributions are correct, the
$p$-values corresponding to the truly null hypotheses will still not be
uniform, and, in general, will have different distributions.

\begin{figure}

\includegraphics{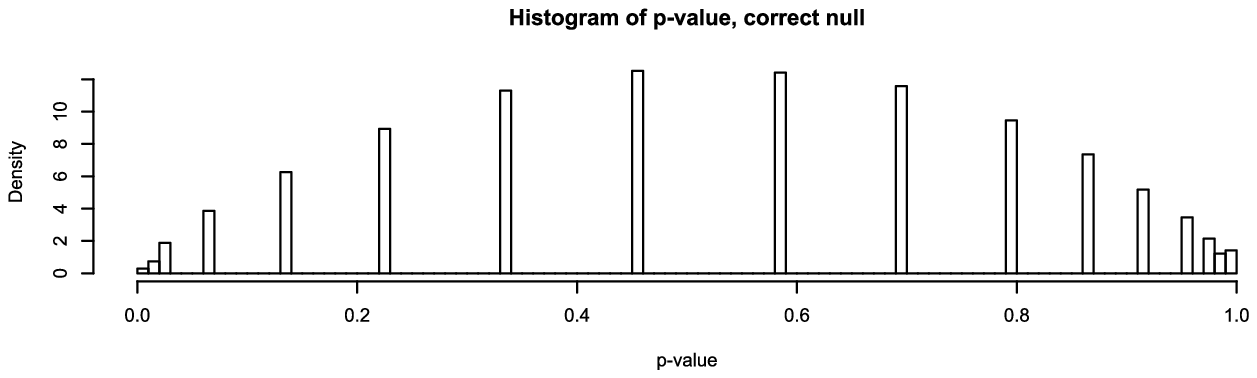}

\caption{$p$-values $p_{i}=F_{i}(x_{i})$, where $x_{i}\sim
F_{i}=\operatorname{Poisson}(10)$.}\label{fig:discrete-correctnull}
\end{figure}

For example, suppose we observe data $x_{i}$, $i=1,\ldots,P$, and we
think that each $x_{i}$ has the same null distribution
$F_{i}=\operatorname{Poisson}(10)$. We can form $p$-values $p_{i}=F_{i}
(x_{i})$ as before. Figure~\ref{fig:discrete-correctnull} shows
that even though our null distributions are correct, the $p$-values are
far from $\operatorname{Unif}(0,1)$. Furthermore, if the null distributions
$F_{i}$ are $\operatorname{Poisson}(\mu_{i})$ with $\mu
_{i}$ varying across $i$, then it is not hard to see that the $p_{i}$
will have different null distributions. Checking the uniformity of the
$p$-values does not tell us if our null distribution is correct or
wrong, and it is not clear how to transform the $p_{i}$ to be uniform.
Because the $p$-values are not uniform under the correct null, we
cannot use the uniformity of the $p$-values to check our nulls. And
since each $p$-value can have a different null distribution even when
our model is correct, it makes little sense to model the $p$-values as
having the same null or marginal distributions. This means that we
cannot use existing methods for estimating empirical nulls and
computing $\mathit{fdr}$'s on discrete data.

%
%f1 ###

%s2.3 ###
\subsection{Randomized $p$-values}

One way to fix this problem is to randomize the $p$-values to make them
continuous. Randomized $p$-values are familiar from classical
hypothesis testing [\citet{Lehmann2005}], and have long been used in the
forecasting literature to assess predictive distributions for discrete
data [\citet{Brockwell2007}, \citet{Czado2009}, \citet{Kulinskaya2009}] recently
used randomized $p$-values to construct versions of the Bonferroni and
Benjamini--Hochberg multiple\vadjust{\goodbreak} testing procedures for discrete data. Their
approach, however, has drawbacks that make it unsuitable for our
purposes. It offers no way to check the nulls, to fit an empirical
null, or to use existing continuous $\mathit{fdr}$ methods. More seriously, it
produces a ``probability of rejection'' for each case, not a false
discovery rate, and is too computationally expensive to apply to even
moderately large data sets.

We propose using existing continuous false discovery rate methods on
randomized $p$-values. Let
%
%e2 ###
\begin{eqnarray}
r_{i} & = & F_{i}^{-}(x_{i})+U_{i}
\bigl(F(x_{i})-F_{i}^{-}(x_{i})\bigr)
\nonumber
\\[-8pt]
\\[-8pt]
\nonumber
& =& P_{F_{i}}(X<x_{i})+U_{i}P_{F_{i}}(X=x_{i}
),\label{eq:rand-pval}
\end{eqnarray}
where $F_{i}^{-}=P(X_{i}<x_{i})$ denotes the left-limit
function of the cdf $F_{i}$, $U_{i}$ are i.i.d. $\operatorname{Unif}(0,1)$
independent of all the $x_{i}$, and $P_{F_{i}}$ denotes probability
under $X\sim F_{i}$. In other words, we use $r_{i}\sim \operatorname{Unif}
(F_{i}^{-}(x_{i}),F_{i}(x_{i}))$ instead of
$p_{i}=F_{i}(x_{i})$.

The key property of $r_{i}$ is that if our null distribution $F_{i}$ is
correct, then $r_{i}\sim \operatorname{Unif}(0,1)$ under the null. This
modification (of $p_i$ to $r_i$) allows us to apply continuous $\mathit{fdr}$
methods to the $r_{i}$. Theorem~\ref{thm:uniformity} makes this
property more precise: The closer $r_{i}$ is to uniform, the closer our
true null distribution is to the assumed null~$F_{i}$, and vice versa.
The theorem (proved in the \hyperref[app]{Appendix}) also holds for the nonrandom
discrete $p$-value functions proposed by \citet{Czado2009}, which can
be used instead of our randomized $p$-values in everything that
follows.
\begin{thmn}
\label{thm:uniformity}Let $x$ be a discrete random variable, $F$
be our predicted distribution for $x$, and $G$ be the true distribution
of $x$. Let $r=F^{-}(x)+U(F(x)-F^{-}
(x))$
be our constructed randomized $p$-value, with density $h(r)$,
cdf $H(r)$, and let $h_{\mathrm{unif}}(t)=1$,
$H_{\mathrm{unif}}(t)=t$
be the uniform density and cdf.

Then
\begin{eqnarray*}
D_{\mathrm{KL}}(H_{\mathrm{unif}}\|H) & = & D_{\mathrm{KL}}(G\|F),\\
D_{\mathrm{KL}}(H\|H_{\mathrm{unif}}) & = & D_{\mathrm{KL}}(F\|G),\\
\sup_{r\in[0,1]}|H(r)-H_{\mathrm{unif}}(r
)| & = & \sup_{x}|F(x)-G(x)|,
\end{eqnarray*}
where for two distribution functions\vspace*{1pt} $P$ and $Q$, $D_{\mathrm{KL}}(P\|
Q)=\int\log(\frac{dP}{dQ})\,dP$
is the Kullback--Liebler divergence. In particular, $r\sim \operatorname{Unif}
(0,1)$
if and only if $F=G$.
\end{thmn}

\begin{figure}

\includegraphics{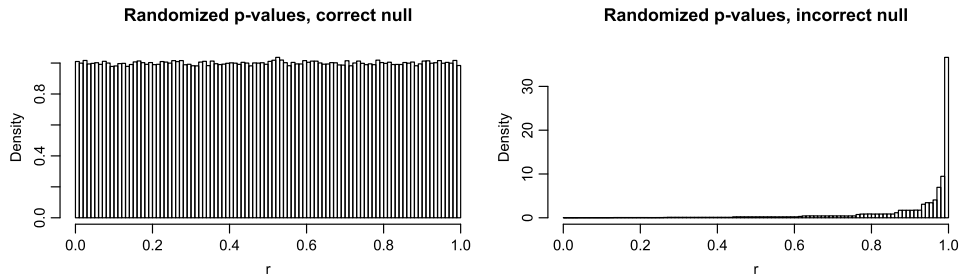}%

\caption{Histograms of randomized $p$-values $r_{i}$ under the correct
$\operatorname{Poisson}(10)$
null (left) and the incorrect $\operatorname{Poisson}(5)$ null (right).
The Kolmogorov distance between the distribution of~$r_{i}$ under
the incorrect null and the uniform distribution is $0.6464$, exactly
the Kolmogorov distance between $\operatorname{Poisson}(5)$ and
$\operatorname{Poisson}(10)$.
The distance from the empirical cdf of the realized $r_{i}$ in the
histogram to the uniform distribution is $0.6465$, which is different
only because of the randomness in $x$ and
$r$.}\label{fig:right-and-wrong}
\end{figure}

Theorem~\ref{thm:uniformity} says that if our null distribution $F_{i}$
is correct, then $r_{i}$ is uniform under the null. Moreover, if our
null distribution is close to the true null in the Kullback--Liebler or
Kolmogorov distance, then $r$ is close to uniform in the same sense
under the null. Consider our previous example, where $x_{i}\sim
\operatorname{Poisson}(10)$. Figure~\ref{fig:right-and-wrong} shows that
$r_{i}$ are uniform if we use the correct\vadjust{\goodbreak} $\operatorname{Poisson}(10)$
null. If we use the wrong null, $\operatorname{Poisson}(5)$, then $r_{i}$
are clearly not uniform. The distance between the distribution of
$r_{i}$ and the uniform distribution is exactly the distance between
the assumed null $\operatorname{Poisson}(5)$
and the correct null $\operatorname{Poisson}(10)$.
%
%f2 ###

Theorem~\ref{thm:uniformity} lets us check our null distributions, fit
empirical null distributions, and estimate false discovery rates using
tools developed for continuous data. Consider the first problem,
checking the null distributions. We know that most $x_{i}$ each come
from their null distribution, and that if we have assumed the correct
null distributions, $r_{i}\sim \operatorname{Unif}(0,1)$ under the null. We
can check for systematic departures from the assumed null distributions
by assessing the $r_{i}$ histogram just as we checked our nulls using
the $p$-value histogram for continuous data, using any model assessment
tool from the continuous $\mathit{fdr}$ literature.

Next, consider estimating an empirical null distribution. We can use
continuous empirical null methods to fit a null distribution $H$ to
$r_{i}$. Just as in the continuous case, we can then use $H$ to fix our
null distributions, changing $F_{i}$ to $\tilde{F}_{i}=H\circ F_{i}$,
and substituting $\tilde{F}$ in place of $F$ in (\ref{eq:rand-pval}) to
make new randomized $p$-values $\tilde{r}_{i}$. Theorem \ref
{thm:uniformity} says that if $\tilde{r}_{i}$ is approximately uniform,
$\tilde{F}_{i}$ is close to the true null distribution.

Finally, consider estimating $\mathit{fdr}$. Using Bayes' rule, we can write
\[
\mathit{fdr}(x_{i})=\frac{P(H_{i0})P_{\mathrm{null}}
(x_{i})}{P_{\mathrm{marg}}(x_{i})},
\]
where $P_{\mathrm{null}}$ and $P_{\mathrm{marg}}$ are the null and marginal distributions
of $x_{i}$. Rewriting in terms of $\tilde{r}_{i}$, this is,
%
%e3 ###
\begin{equation}
\mathit{fdr}(x_{i})=\frac{P(H_{i0})P_{\mathrm{null}}(\tilde
{r}_{i}\in[\tilde{F}_{i}^{-}(x_{i}),\tilde{F}_{i}
(x_{i})])}{P_{\mathrm{marg}}(\tilde{r}_{i}\in[\tilde
{F}_{i}^{-}(x_{i}),\tilde{F}_{i}(x_{i})
])}.\label{eq:rand-fdr}\vadjust{\goodbreak}
\end{equation}
As before, we bound $P(H_{i0})$ by $1$, and since
$\tilde{r}_{i}$ are uniform under the null,
\[
P_{\mathrm{null}}\bigl(\tilde{r}_{i}\in[\tilde{F}_{i}^{-}(x_{i}
),\tilde{F}_{i}(x_{i})]\bigr)=\tilde{F}_{i}
(x_{i})-\tilde{F}_{i}^{-}(x_{i}).
\]
We can model the $\tilde{r}_{i}$ as having approximately the same
marginal distribution since they are all $\operatorname{Unif}(0,1)$ under
the assumed null distribution.
This lets us use the distribution of $\tilde{r}_{i}$ to estimate the
marginal probability in the denominator of (\ref{eq:rand-fdr}).
Substituting these three values into (\ref{eq:rand-fdr}) gives us an estimated
false discovery rate. Randomization thus lets us translate the three
key steps in a continuous $\mathit{fdr}$ analysis to the discrete setting.

It is important to note that although we use randomized $p$-values, the
variability in the randomization does not significantly affect our
final $\mathit{fdr}$ estimates. Given~$\tilde{F}$, the false discovery rate in
(\ref{eq:rand-fdr}) is a deterministic function of the data~$x$, so the
randomization step affects our $\mathit{fdr}$ estimate only through the
estimated empirical null~$\tilde{F}$ and the marginal distribution of~$\tilde{r}_{i}$.
These quantities depend on the empirical distribution
of all or most of the~$\tilde{r}_{i}$'s, and do not depend strongly on
any individual~$\tilde{r}_{i}$. For large $P$, the empirical
distribution of $\tilde{r}_{i}$ will be close to its true distribution,
which is a~deterministic function of the $x_{i}$'s. Thus, for large
$P$, the variability in the randomization will have little effect on
our $\mathit{fdr}$ estimates. For small $P$, if the extra variability from
randomization is a concern, we can substitute the nonrandom $p$-value
functions proposed by \citet{Czado2009} with essentially no change to
our analysis.

%s3 ###
\section{Modeling sequencing error rates}\label{sec:Modeling-Sequencing-Error}

In this section, we turn to the application of detecting DNA mutations
and present an empirical Bayes model for sequencing error rates.
Mutations appear in the data as unusually high observed error rates, so
detecting mutations accurately requires understanding the normal
variation in error rates. We begin by describing two example data sets
and summarizing the existing approaches. Then, we describe
a~hierarchical model for observed error rates that accounts for sample
effects, genome position, and finite depth. Our model shares
information across positions and samples to estimate error rates and
quantify their variability.

%s3.1 ###
\subsection{Example data sets: Virus and tumor}\label{sub:Example-Data-Sets:}

Our first example is motivated by the problem of detecting rare
mutations in virus and microbial samples. Deep, targeted sequencing has been
used to identify mutations that are carried by a very small proportion
of individuals in the sample. Detecting these rare mutations is
important, because they represent quasispecies that may
expand after vaccine treatment. We use the synthetic DNA admixture data from
\citet{Flaherty2011}, in which a reference and a mutant version of
a synthetic 281 base sequence are mixed at varying ratios. The mutant
differs from the reference at 14 known positions. This data set contains
six samples, 3 of which are 100$\%$ reference, the other 3 contain
a 0.1$\%$\vadjust{\goodbreak} mixture of the mutant sequence. These samples were sequenced
on an Illumina GAIIx platform. The reads were then aligned to the
reference sequence,
yielding nonreference counts (``errors'') $x_{ij}$ and depth
$N_{ij}$ for each position ($i=1,\ldots,P=281$, $j=1,\ldots,S=6$) [see
\citet{Flaherty2011} for more details].
Our goal is to find the mutations, which appear in the data as unusually
large error rates $x_{ij}/N_{ij}$.

Our second example is a comparison of normal and tumor tissue in $S=28$
lymphoma patients, plus tissue from one healthy individual sequenced
twice as a control. A set of regions containing a total of $P=309\mbox{,}474$
genome positions was extracted from each sample and sequenced on the
Illumina GAIIx platform, yielding nonreference counts $x_{ij},y_{ij}$
and depths $N_{ij},M_{ij}$ for the normal and tumor tissues. Our
goal is to find positions that show biologically interesting differences
between the normal and tumor samples, such as positions that are mutated
in the tumor or variant positions in the normal that have seen a loss
of heterozygosity. These appear in the data as significant differences
between the error rates $x_{ij}/N_{ij}$ and $y_{ij}/M_{ij}$.

The two detection problems pose different challenges. Since virus
genomes are short, they can be sequenced to uniformly
high depth. For example, the synthetic virus data from \citet{Flaherty2011}
has depth in the hundreds of thousands. Human tissue, however, is
usually sequenced to a lower, more variable depth. The tumor data has
a median depth of $171$, but the depth varies over five orders of
magnitude, from $0$ to over $100\mbox{,}000$. Discreteness
is thus a more serious problem for the tumor application than it is for
the virus application. The tumor data also exhibits
much more variation in error rates, from less than $0.1\%$ to over
$20\%$, because the human genome is harder to target and map.

Analyzing the virus data is difficult primarily because we are interested
in very rare mutations. A mutation carried by $0.1\%$ of the viruses
may be biologically interesting, but one carried by $0.1\%$ of the
tumor cells is typically less interesting, since biologists usually are
interested in mutations present in a substantial fraction of the tumor
cells. Despite the high sequencing depth, it is difficult to
detect such a small change in base proportions using discrete counts.

%s3.2 ###
\subsection{Existing approaches} \label{sec:existing}

Most current methods for variant detection in sequencing data are
designed to analyze samples of DNA from pure, possibly diploid, cells.
In pure diploid samples, variants are present at levels of either $50\%
$ or $100\%$ of the sample, and are thus much easier to detect than
variants in mixed samples, where they may be present at continuous
fractions. Nearly all existing methods, including the widely used
methods of \citet{Li2008} and \citet{McKenna2010}, rely on sequencing
quality scores from the Illumina platform and mapping quality metrics
to identify and filter out high-error positions. Storing and processing
these quality metrics is computationally intensive, and methods
utilizing these metrics are not portable across experimental platforms.\vadjust{\goodbreak}

\citet{Muralidharan2011} proposed a method to detect single nucleotide
variants in normal diploid DNA. Their method uses a mixture model with
mixture components corresponding to different possible genotypes, and
pools data across samples to estimate the null distribution of
sequencing errors at each position. They showed that this approach,
which avoids using quality metrics, outperforms existing quality metric
based approaches.

A different approach to variant detection was proposed by \citet
{Natsoulis2011}, who use techniques based on domain knowledge, such as
repeat masking (see Section~\ref{sec:Tumor-Data}) and double-strand
confirmation (evidence for the variant must be present in both the
forward and reverse reads covering the position) to identify high-error
positions and eliminate false calls. This method can also be used to
call mutations in tumors using matched normal samples.

Although most current methods for variant detection are designed for
pure diploid samples, a few methods for detecting rare variants in
virus data have recently been proposed. \citet{Hedskog2010} find simple
upper confidence limits for the error rate and use them to test for
variants. \citet{Flaherty2011} use a Beta-Binomial model, that is, less
conservative but much more powerful. Their model for sequencing error
rates is similar in form to ours, but uses a Beta distribution for
error rates that we find does not fit the data. \citet{Hedskog2010} and
\citet{Flaherty2011} also do not account for the effects of sample
preparation on the error rates. Finally, both papers simply use a
Bonferroni bound to avoid multiple testing concerns. This is reasonable
since the data they analyze have only a few hundred positions, but it
makes their methods inapplicable to large genomic regions where
multiple testing is a more serious problem.

%s3.3 ###
\subsection{Sequencing error rate variation}

Sequencing error rates show three types of variation. The first type
of error rate variation comes from finite depth. Consider a nonmutated
position, where all nonreference counts are truly errors. Given the
depth and an error rate, we can model the nonreference counts as
binomial,
%
%e4 ###
\begin{equation}
x\sim \operatorname{Binomial}(N,p).\label{eq:binomial}
\end{equation}
Because $N$ is finite, the observed error rate $x/N$ will vary around
the true error rate $p$. This type of variation is easily handled by
the binomial model.

%f3 ###
\begin{figure}

\includegraphics{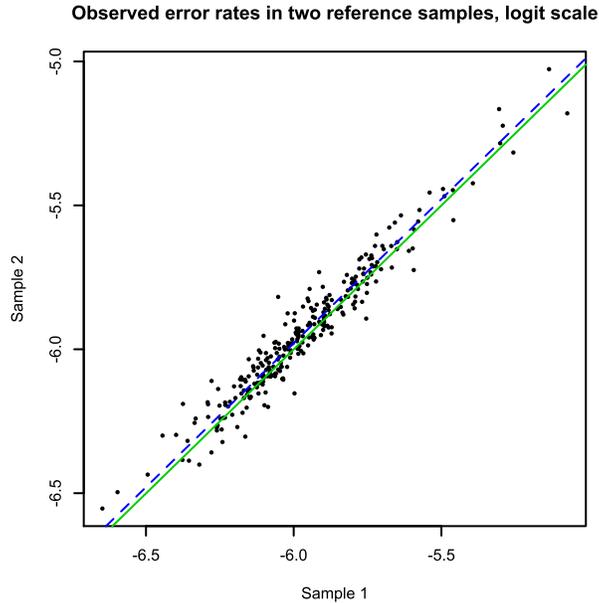}

\caption{Observed error rates $\logit(x/N)$ for two reference samples
in the virus data. The solid green line is the $x=y$ diagonal. The
error rates for the second sample are slightly but significantly biased
upward. The dashed blue line shows the diagonal shifted to account
for this bias.}\label{fig:log-log-virus}
\end{figure}

The second type of variation is positional: as shown by \citet
{Muralidharan2011} and \citet{Flaherty2011}, different positions in the
genome have different error rates. This means that each position
has its own error rate $p$ in our binomial model (\ref{eq:binomial}).
Suppose we have extremely large depth, so that the binomial variation
in the observed error rate $x/N$ is negligible. A large observed error
rate at a given position is still not enough to report a mutation,
because that position may simply be noisy. We can account for the
positional variation in error rates by aggregating data across samples
to estimate the baseline sequencing error $p$ at each position.

The last type of variation is variation across samples. Small
differences in sample preparation and sequencing, such as the sample's
lane assignment on the Illumina chip, can create differences in the
sequencing error rate at each position, even when the sample contains
no mutations. For example, suppose that we have extremely large depth,
that we have estimated the positional error rate $p$ perfectly, and
that we observe an error rate $x/N$, that is, higher than $p$. We
still cannot conclude that the position is mutated, because the difference
between $x/N$ and $p$ may be due to sample preparation. We can account
for cross-sample variation by aggregating data across positions to
estimate sample effects.

Figure~\ref{fig:log-log-virus} illustrates these three sources of variation.
It plots, on the logit scale, observed error rates $x/N$ for two
reference samples from the synthetic data of \citet{Flaherty2011}.
Each point in the plot represents a position. There are no mutant
positions, so all points represent null observed error rates. The
figure shows that error rates in the two samples are highly correlated
and depend strongly on genome position. The binomial variation due
to finite depth causes some of the spread around the diagonal. Sample
variation also causes spread around the diagonal, as well as a systematic
bias-error rates for the second sample are slightly but significantly
higher than error rates in the first. These two samples were actually
sequenced in the same lane; we observed stronger sample effects when
comparing data from different lanes. We also saw similar behavior on
the tumor data.\vspace*{-3pt}

%f4 ###
\begin{figure}

\includegraphics{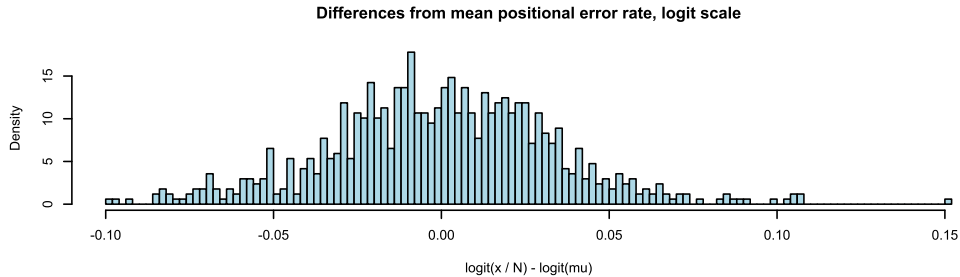}%
\vspace*{-3pt}
\caption{Differences between observed error rate $x/N$ and positional error
rate $\mu$ for the three reference samples, on the logit scale. The
mean error rate $\mu$ was estimated by averaging $\logit(x/N)$
for each position over the reference samples.}\label{fig:logitnormal-diffs}\vspace*{-3pt}
\end{figure}

%s3.4 ###
\subsection{Modeling the variation} \label{sec:modelvariation}

Figure~\ref{fig:log-log-virus} also suggests a model for sequencing
error rates---when plotted on the logit scale, the error rates are
dispersed evenly around a shifted diagonal. Figure~\ref{fig:logitnormal-diffs}
shows that the dispersion in error rates is roughly normal. Accordingly,
we can model the logit sequencing error rate in each sample as a sum
of a positional error rate, sample bias, and normally distributed
sample noise. Given the error rate, we observe binomial counts. This
model makes sense biologically: sample preparation for these two data
sets includes PCR amplification,
an exponential process, so it is plausible that differences in sample
preparation produce additive effects on the logit scale.

This formulation yields the following hierarchical model for the
unmatched mutation detection problem such as in the virus application:
%
%e5 ###
\begin{eqnarray}\label{errmodel}
\logit p_{ij} & \sim& \mathcal{N}(\logit\mu_{i}+\delta_{j},\sigma
_{j}^{2}),
\nonumber
\\[-9pt]
\\[-9pt]
\nonumber
x_{ij}|p_{ij} & \sim& \operatorname{Binomial}(N_{ij},p_{ij}),
\end{eqnarray}
where $\mu_{i}$ is the positional error rate, $\delta_{j}$ is a
sample-specific error rate bias (constant across positions), and $\sigma_{j}$
measures the sample specific noise in error rates. Fitting~$\mu$,
$\delta$, and $\sigma$ provides information on the positional error rates,
sample biases, and cross-sample variability in our data.

This model allows us to test whether an observed error rate is unusual
enough to be a mutation. For example, consider applying the model
to the virus data. Once we fit the parameters, as described in Section
\ref{sec:fit}, the model gives a null distribution for the observed
error rate at each position. We can then compare the observed error
rates for each position in a clinical sample to its null distribution
and use the false discovery rate methods from Section~\ref{sec:Multiple-Testing-Tools}
to find mutated positions.\vadjust{\goodbreak}

Next, consider tumor data with matched normals.
We model the normal tissue error rates as in (\ref{errmodel}), and
introduce extra
parameters to account for additional error rate variation between
normal and tumor tissue from the same patient:
\begin{eqnarray*}
\logit p_{ij} & \sim& \mathcal{N}(\logit\mu_{i}+\delta_{j},\sigma
_{j}^{2}),\\
\logit q_{ij}|p_{ij} & \sim& \mathcal{N}(\logit p_{ij}+\eta
_{j},\tau_{j}^{2}),\\
x_{ij}|p_{ij},q_{ij} & \sim& \operatorname{Binomial}(N_{ij},p_{ij}),\\
y_{ij}|p_{ij},q_{ij} & \sim& \operatorname{Binomial}(N_{ij},q_{ij}),
\end{eqnarray*}
where $p_{ij}$, $q_{ij}$ are the normal and tumor error rates,
respectively; $\delta_{j},\eta_{j}$ are sample effects, $\sigma_{j}$
is the noise variance for the normal tissue, and $\tau_{j}$ is the
noise variance for the difference between tumor and normal tissue.
After fitting the parameters as described in Section~\ref{sec:fit}, we
use this model to find the conditional null distribution
for the tumor error rates, given the observed normal error
rates. That is, we use the model to find null distributions for
\[
\frac{y_{ij}}{M_{ij}}\bigg|\frac{x_{ij}}{N_{ij}},
\]
and then use the false discovery rate approach in Section \ref
{sec:Multiple-Testing-Tools}
to find mutated positions.

The logit-normal model naturally handles the discreteness and wide
range of depths in our data. It separates the observed error rate
variation into depth, positional variation, and sample effects, and
combines the different sources of variation to give the appropriate
null distribution in each case.
%The model is a convenient way to share information across positions
%and across samples. We share information across samples to estimate
%the positional error rates $\mu$, and across positions within a sample
%to estimate the sample biases $\delta,\eta$ and variability $\sigma,

%s3.4.1 ###
\subsubsection{Fitting\label{sec:fit}}

The best way to fit our model will depend on the data set, so we will
discuss the fitting in only general terms.

Estimating $\mu,\delta$, and $\eta$ is usually straightforward. For
example, in the virus data set, we use the median of the observed error
rates for each position over all of the reference samples to estimate
$\mu$, then estimate $\delta$ using all of the positions in each
sample,
\[
\hat{\delta}_{j}=\operatorname{median} \biggl(\logit\frac{x_{ij}}{N_{ij}}-\logit
\hat{\mu}_{i}\biggr).
\]
Similar ideas can also be applied to estimate these parameters for the
tumor data.

Estimating the sample error rate variances $\sigma$ and $\tau$ can be
more difficult. The simplest and fastest approach is to use the method
of moments as an approximate version of maximum likelihood. This works
well if depths are large, as in the virus data. If depths are small, as
in the tumor data, the method of moments works badly and it is better
to use the maximum likelihood.\vadjust{\goodbreak}

The tumor data also has extra sources of variability, which we discuss
briefly to illustrate how our method can be adapted to the specific
characteristics of a data set. Because of genetic variation between
people, not all normal samples have the same base at each position. For
example, at single nucleotide polymorphic positions (SNPs),
heterozygous samples have an observed ``error rate'' close to $0.5$
against the reference genome, while homozygous samples have an observed
error rate close to $0$. We account for SNP positions by using a simple
mixture model to genotype the samples and estimating $\mu_i$ separately
for each genotype. We also increase $\sigma_j ^ 2$ for positions with
multiple genotypes to account for the extra uncertainty due to possibly
incorrect genotyping.

Another source of extra variability comes from the technology used to
generate our data set: The 309,474 genome positions are regions of the
genome that have been targeted by primers and amplified. We observe
empirically that regions treated with some primers have more variable
error rates across samples. These regions can be identified using extra
data generated by the sequencer. We account for this extra variability
by fitting different error variances $\sigma_j$ and $\tau_j$ for each
genomic region, and using a high quantile of the region-wise
variabilities as our $\sigma_j$.

The logit-normal prior for $p$ makes it difficult to calculate the
marginal distributions of counts, find predictive distributions, and
fit $\sigma,\tau$ by maximum likelihood. We approximate the
logit-normal distribution with a Beta distribution. If
\[
p\sim \operatorname{Beta}\biggl(\frac{1}{\sigma^{2}(1-\mu)},\frac{1}{\sigma
^{2}\mu}\biggr),
\]
then it is easy to show using Stirling's formula that $\logit p$ has
approximate mean $\logit\mu$, variance $\sigma^{2}$, skewness
\[
\sigma\bigl(\mu^{3}-(1-\mu)^{3}\bigr),
\]
and excess kurtosis
\[
2\sigma^{2}\bigl(\mu^{4}+(1-\mu)^{4}\bigr).
\]
If $\sigma$ is small and $\mu$ is close to $0$ or $1$, as they are in
our data, then $\logit p$ is approximately $\mathcal{N}(
\logit\mu,\sigma^{2})$. This Beta approximation
makes it much easier to calculate marginal
and posterior distributions.

%f5 ###
\begin{figure}

\includegraphics{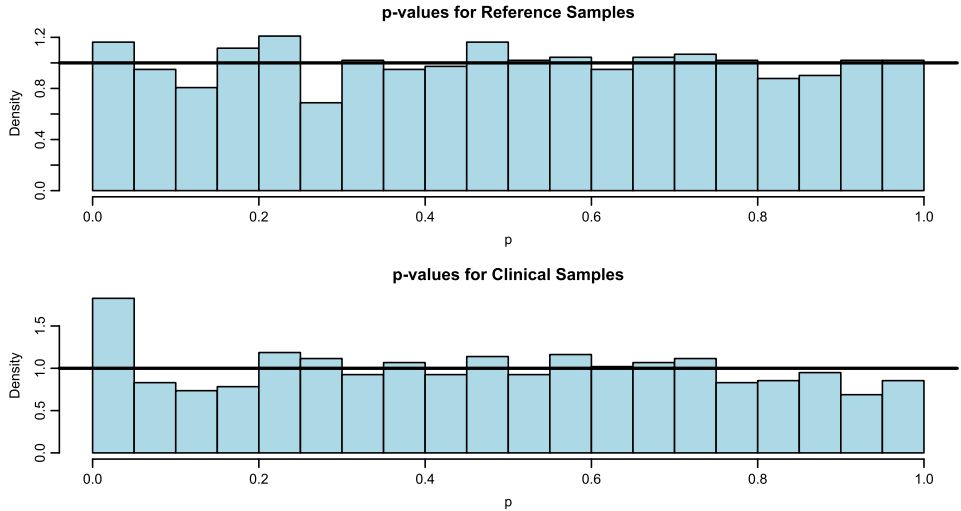}

\caption{Histogram of $p$-values for the virus data, reference samples
(top plot) and clinical samples (bottom plot).}\label{fig:dilution-pval}
\end{figure}

%t1 ###
\begin{table}[b]
\caption{Detection results on clinical samples
of the synthetic virus data}\label{tab:dilution-results}
\begin{tabular*}{\textwidth}{@{\extracolsep{\fill}}lccc@{}}
\hline
& \textbf{Our method,} & \textbf{Our method,} &\\
& $\bolds{\hat{\mathit{fdr}}\leq0.1}$ &$\bolds{\hat{\mathit{fdr}}\leq0.01}$ & \textbf{Flaherty et al.}\\
\hline
True positives (of 42) & 42\phantom{0} & 39\phantom{00} & 42\phantom{0}\\
False positives & \phantom{.}1 & 0\phantom{0} & 10\phantom{0}\\[3pt]
{Power} & {100\%} & {93\%} & {100\%}\\
{False positive rate} & \phantom{0000.}{2.32\%} & \phantom{0}{0\%} &
\phantom{000.}{19.23\%}\\
\hline
\end{tabular*}
\end{table}

%s4 ###
\section{Results}\label{sec:Results}

%s4.1 ###
\subsection{Virus data}

We first tested our method by applying it to the virus data, described
in Section~\ref{sub:Example-Data-Sets:}. In this synthetic data, we
know the locations of the 14 variant positions, and we know that the
mutant base is present in $0.1\%$ of the viruses in each case. We did
not use any information about the mutations' location or prevalence
when fitting our model. Thus, we can use this data to evaluate our method's
power and specificity.\vadjust{\goodbreak}

Our model fits the data reasonably well. Figure~\ref{fig:dilution-pval}
shows the $p$-values histograms for the reference and clinical samples;
randomization is unnecessary since the depth is so high
(the median depth is $775\mbox{,}681$, and $95\%$ of positions have depth
between $271\mbox{,}192$ and $1\mbox{,}689\mbox{,}977$). The $p$-values are fairly uniform
for the reference samples, and also uniform in the clinical samples
except for a spike near $0$ that indicates that some positions are
truly nonnull. Since our null distributions fit the data accurately
enough, we did not need to estimate an empirical null. We used the
log-spline $\mathit{fdr}$ estimation method proposed by \citet{Efron2004a}
to estimate the false discovery rate $\hat{\mathit{fdr}}_{ij}$ for each position
in each sample. Finally, we declared any position with $\hat{\mathit{fdr}}$
less than a given threshold to be a mutation.

Table~\ref{tab:dilution-results} compares our results to the method of
\citet{Flaherty2011}. Our method produces fewer false\vadjust{\goodbreak} discoveries while
maintaining excellent power. If we use an $\hat{\mathit{fdr}}$ threshold of $10\%
$, our method detects all $42$ mutations (14~in each clinical sample)
and makes $1$ false discovery, for a false positive rate of $2.3\%$. A
more stringent $\mathit{fdr}$ threshold of $1\%$ eliminates all false
discoveries, at the cost of missing~$3$ mutations. Our method's high
power and low false discovery rate is especially notable given that the
mutation is only present at $0.1\%$ within the sample.\vspace*{-3pt}
%

%s4.2 ###
\subsection{Tumor data}\label{sec:Tumor-Data}

Next, we applied our method to the tumor data, also described in
Section~\ref{sub:Example-Data-Sets:}. Our model fits the data
relatively well, but not as accurately as it fits the virus data. We
can assess the model by examining the last sample pair, which actually
consists of a healthy person's normal tissue that was sequenced twice
as though it were normal and tumor tissue.

\begin{figure}[b]
\vspace*{-3pt}
\includegraphics{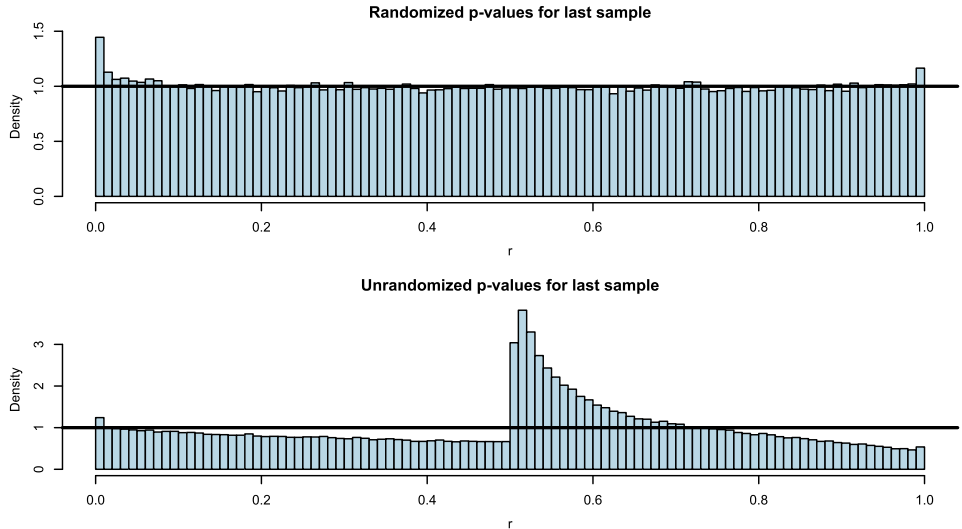}%
\vspace*{-3pt}
\caption{Randomized $r_{ij}$ (top plot) and unrandomized $p_{ij}$
(bottom plot) values for the last normal tumor pair (actually
the same normal tissue sequenced twice).}\label{fig:rpval-lastpair}
\end{figure}

Figure~\ref{fig:rpval-lastpair} shows the histogram of randomized and
unrandomized $p$-values for the last sample pair. The randomized
$p$-values $r_{ij}$ are uniform through most of the unit interval,
indicating that most of our fitted null distributions are close to the
true null distributions. In contrast, the unrandomized $p$-value
histogram tells us next to nothing about our null distributions.
%
%f6 ###

Our null distributions do not give a perfect fit: the $r_{ij}$ appear
to be enriched near~$0$ and $1$, so if we thought the null were
uniform, our false discovery rates would be misleadingly small near $0$
and $1$. Empirical nulls are not very helpful here, because they are
fit to the center of the distribution rather than the tails. Inspecting
the sample reveals that the null distribution is enriched near $0$ and
$1$ because the error rates $p$ and $q$ are more variable very close to
$0$ and $1$ than our normal model predicts. We will discuss this issue
a bit more later.\vadjust{\goodbreak}

\begin{figure}

\includegraphics{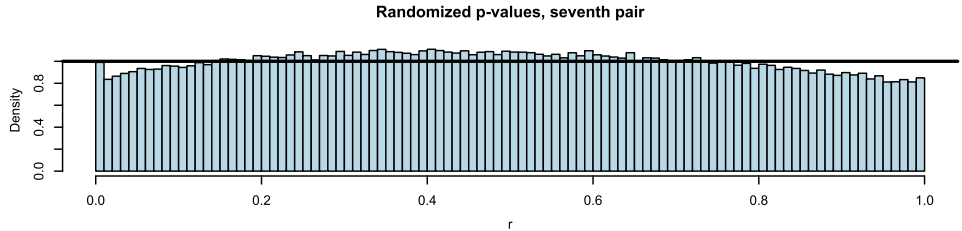}

\caption{Randomized $p$-values
$r_{i}$ for the seventh normal tumor pair. This sample had the least
uniform $r_{i}$.}\label{fig:rpval-worstpair}
\end{figure}

Although our null distributions are mostly correct for the last sample,
they are not as good on some other samples. Figure \ref
{fig:rpval-worstpair} shows the randomized $p$-value histogram for the
seventh sample pair, which shows the most deviation from uniformity.
The underdispersion in Figure~\ref{fig:rpval-worstpair} means that our
null distributions are systematically too wide on that sample.
%
%f7 ###

%
%f8 ###
\begin{figure}[b]

\includegraphics{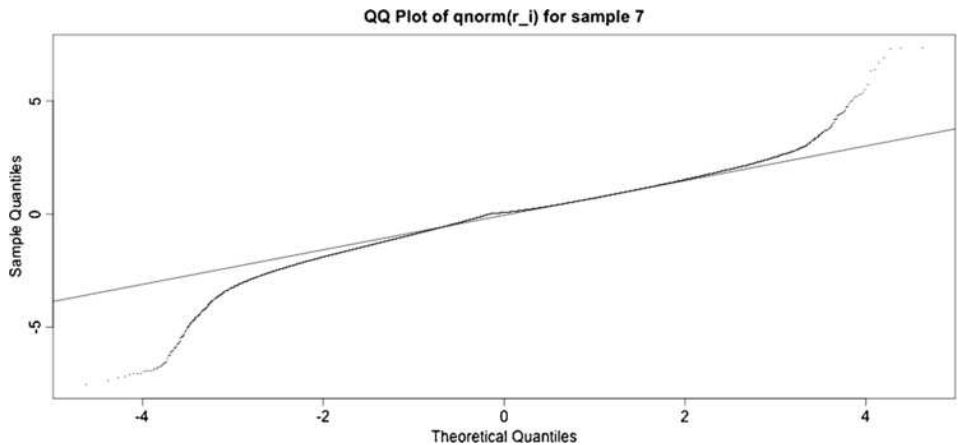}

\caption{Normal QQ plot of $z_{ij}=\Phi^{-1}(r_{ij})$ for sample
$7$.}\label{fig:Normal-QQ-plot}
\end{figure}

\begin{figure}

\includegraphics{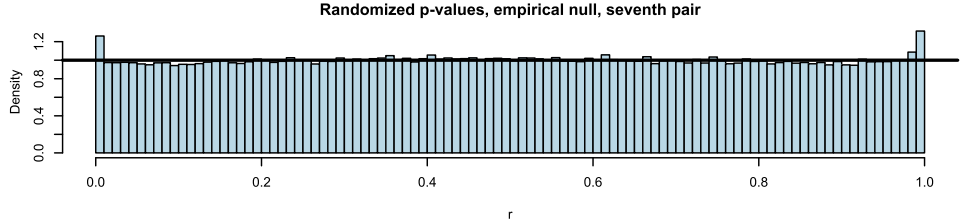}

\caption{Empirical null randomized $p$-values $\tilde{r}_{i}$ for the seventh
normal tumor pair. The empirical null yields much more uniform $p$-values
(compare to Figure \protect\ref{fig:rpval-worstpair}).}\label{fig:rpval-worstpair-cor}
\end{figure}

We fit empirical nulls to correct our null distributions. Figure \ref
{fig:Normal-QQ-plot} shows a~normal quantile--quantile plot of
randomized $p$-values for sample $7$, transformed to the normal scale
by $z_{ij}=\Phi^{-1}(r_{ij})$. The QQ plot is straight
through the bulk of the data, indicating that our null can be corrected
by centering and scaling on the normal scale. Our corrected null will
still be too light-tailed in the far tails, but, as for the last
sample, these points correspond to very small changes in error rate
very close to $0$ and $1$, which we will discuss later. Accordingly, we
used the median and a robust estimator of scale [$S_n$, described by
\citet{Rousseeuw1993}] on $z_{ij}$ to estimate a location and scale for
our empirical null in each sample. Figure~\ref{fig:rpval-worstpair-cor}
shows that this yielded much more uniform randomized $p$-values.

Finally, we estimated the density of the empirical null adjusted
randomized $p$-values using a log-spline. We then estimated the $\mathit{fdr}$.
To ease computation, we approximated the $\mathit{fdr}$ expression in equation
(\ref{eq:rand-fdr}). Instead of estimating
\[
P_{\mathrm{marg}}\bigl(\tilde{r}_{i}\in[\tilde{F}_{i}^{-}(x_{i}
),\tilde{F}_{i}(x_{i})]\bigr),
\]
we fit $f_{\mathrm{marg}}$ and used the approximation
%
%e7 ###
%e6 ###
\begin{eqnarray}
& & P_{\mathrm{marg}}\bigl(\tilde{r}_{i}\in[\tilde{F}_{i}^{-}
(x_{i}),\tilde{F}_{i}(x_{i})]\bigr)\\
&&\qquad \approx \hat{f}_{\mathrm{marg}}\bigl(\tfrac{1}{2}\bigl(\tilde{F}_{i}^{-}
(x_{i})+\tilde{F}_{i}(x_{i})\bigr)\bigr)\bigl(\tilde
{F}_{i}(x_{i})-\tilde{F}_{i}^{-}(x_{i})\bigr).
\label{fmarg}
\end{eqnarray}
Substituting (\ref{fmarg}) into (\ref{eq:rand-fdr}) yields an estimate
of the false discovery rate~$\hat{\mathit{fdr}}_{ij}$ for each position in each sample.

As mentioned, many positions had a low $\hat{\mathit{fdr}}$ while being
biologically uninteresting due to the heavier tail of the null $p$-value
distribution around 0 and 1. Our model looks at differences between
normal and tumor error rates on the logit scale, which exaggerates
differences near~$0$ and $1$; for example, on the logit scale, $0.001$
and $0.003$ are as far from each other as $0.5$ and $0.75$. Such small
changes near~$0$ and $1$ are also more likely to be false positives,
since null error rates are more variable very near~$0$ and $1$ than our
model predicts. Even if they were real, mutations present at such small
fractions in tumor tissue are too rare to be biologically interesting.
For most tumor analysis scenarios, we want to find mutations that are
present in a fairly large fraction of the cells in the tumor tissue,
with the prevalence threshold determined
by the biologist.

To find such mutations, we estimated the change in error rate at each
position for each sample using a very simple {}``spike and slab''
model. We supposed that either the normal and tumor error rates were
the same, or they were different, in which case we knew nothing about
either. Under this model, the expected error rate difference given the
data is
\begin{eqnarray*}
\Delta_{ij} & = & E(q_{ij}-p_{ij}|x,y)\\ &
= & P(q_{ij}\neq p_{ij}|x,y)\biggl(\frac{y_{ij}}{M_{ij}}-\frac
{x_{ij}}{N_{ij}}\biggr),
\end{eqnarray*}
which we can estimate by
\[
\hat{\Delta}_{ij}=\hat{\mathit{fdr}}_{ij}\biggl(\frac{y_{ij}}{M_{ij}}-\frac
{x_{ij}}{N_{ij}}\biggr).
\]
We required a position to have a large $\hat{\Delta}$ ($|\hat
{\Delta}|\geq0.25$) as well as a low $\hat{\mathit{fdr}}$ $(\hat
{\mathit{fdr}}\leq0.1)$ to
be called a biologically interesting mutation.

Thresholding for both false discovery rate and estimated effect size
yielded 427 mutation calls on the clinical samples. Assessing these
calls is difficult. Unlike for the synthetic data, we do not know
which positions are truly mutated or null for the tumor data. Since
all putative mutations in the tumor samples are new changes, and would
be unique to each sample, we cannot assess our mutation calls using
databases of known variants. Also, targeted deep resequencing is
currently the best technology for variant detection, so, short of
resequencing the entire genomic region at even higher depth, we cannot
use some other gold-standard experimental method to validate our calls.

We therefore use a simple domain-knowledge based proxy, enrichment
in repetitive regions, as a crude check that our method gives useful
results. Repetitive regions are segments of DNA that repeat themselves
with high sequence similarity at multiple places in the genome. They
confuse the DNA targeting, extraction, and mapping steps in the
experiment, and have been a major source of false calls for previous
variant detection methods. Because of this, most existing variant
detection methods use repeat detection algorithms to find repetitive
regions, and then use the output of these algorithms to refine their
calls. The most common approach has been to simply ignore calls in
regions that are designated as repetitive, since otherwise the calls
would be dominated by false calls in these regions.

Masking repetitive regions has some disadvantages. First, different
repeat detection algorithms often disagree, so the choice of repeat
detection method and associated parameters can substantially impact the
final list of calls. Second, many functional areas of the genome, such
as exons, contain repeated genetic material. For
example, roughly $8.5\%$ of our tumor data, which consists almost
entirely of exons, lie in repetitive regions (the exact percentage
depends on the repeat detector and parameters used). If we simply
ignore mutation calls in repetitive regions, we
may miss important mutations in functional regions.

Our approach does not rely on any information about whether a position
lies in a repetitive region. The high error rates in repetitive regions
are reproducible across samples, and thus by modeling the error rate as
a function of genome position, we can account for the higher error
rates in repetitive regions without using any explicit information
about repetitiveness.

Of the $427$ mutations found in the tumor data by our method, $95$
($22.1\%$) lie in repetitive regions. In comparison, \citet{Natsoulis2011}
make $1305$ calls before\vadjust{\goodbreak} their final repeat masking step, $470$
($36\%$) of which are in repetitive regions. Although our calls are
somewhat enriched in repetitive regions, they are less enriched than
the calls made by \citet{Natsoulis2011} before repeat masking, despite
not using any domain knowledge explicitly. This is a rough indication
that our positional error-rate model is estimating higher error rates
in repetitive regions.

Our method makes more calls than \citet{Natsoulis2011} in low depth
regions. We make a $233$ gain of allele calls, $47$ ($20.1\%$) of which
are in repetitive regions. Of the $186$ calls we make outside of
repetitive regions, $103$ are among the $165$ gain of allele calls made
by \citet{Natsoulis2011}. Nearly half of the $83$ calls made by our
method outside repetitive regions and not made by \citet{Natsoulis2011}
are in low depth regions of the genome. We would like to think that
this indicates our method
is able to achieve higher power in low depth regions by pooling data
across samples to estimate the null distribution of the error rates.
We cannot know the truth, however, without a rigorous validation
experiment.\vspace*{-3pt}

%s4.3 ###
\subsection{Summary}

In this paper, we have shown that empirical Bayes ideas can be usefully
applied to detect mutations in high throughput sequencing data from
mixed DNA samples. We used a hierarchical model to account for
different sources of variation in sequencing error rates. This model
let us weigh the different sources against one another, and naturally
accommodates the discreteness and depth variation in our data. We also
adapted continuous $\mathit{fdr}$ methods to discrete data using a simple
randomization scheme. Combining the new multiple testing methods with
the empirical null distributions for sequencing error rates yielded a
powerful, statistically sound way to detect mutations in mixed samples.\vspace*{-3pt}

\begin{appendix}
\section*{Appendix}\label{app}
We prove Theorem~\ref{thm:uniformity}, which justifies the use of
randomized $p$-values.
From the construction of $r$, we have that
\[
r|x\sim \operatorname{Unif}(F^{-}(x),F(x)).
\]
Thus, the unconditional density of $r$ is
\[
h(r)=\sum_{x}\frac{P_{G}(x)}{P_{F}(x
)}I_{r\in[F^{-}(x),F(x)]},
\]
where $P_{F}$ and $P_{G}$ denote probability under $F$ and $G$
respectively. This means that
\begin{eqnarray*}
D_{\mathrm{KL}}(H_{\mathrm{unif}}\|hH) & =&\int\log\biggl(\frac
{h(r)}{h_{\mathrm{unif}}(r)}\biggr)h(r)\,dr\\
&=&\sum_{x} \int_{F^-(x)}^{F(x)}\frac{P_G(x)}{P_F(x)}\log\biggl[\frac
{P_G(x)}{P_F(x)}\biggr]\,dr\\
& =&\sum_{x}P_{G}(x)\log\frac{P_{G}(x)}{P_{F}
(x)}\\
& =&D_{\mathrm{KL}}(G\|F).
\end{eqnarray*}
The other Kullback--Liebler equality is proved similarly.

For the Kolmogorov distance, note that the cdf of $r$, $H$, is piecewise
linear, and the uniform cdf $H_{\mathrm{unif}}(r)=r$ is also linear.
This means that $|H-H_{\mathrm{unif}}|$ reaches its maximum at
one of the knots of $H$, and these are~$0$, $1$, and $F(x)$
for all possible values of $x$. Since $H(0)=H_{\mathrm{unif}}
(0)=0$
and $H(1)=H_{\mathrm{unif}}(1)=1$, the maximum has to
occur at some $F(x)$. At these points, though,
\begin{eqnarray*}
H(F(x)) & =&P_{G}\bigl(r\leq F(x)
\bigr)\\
& =&\sum_{x^{*}\leq x}\bigl(F(x^{*})-F^{-}(x^{*}
)\bigr)\frac{P_{G}(x^{*})}{P_{F}(x^{*})}\\
& =&G(x)
\end{eqnarray*}
so
\begin{eqnarray*}
\sup_{r}|H(r)-H_{\mathrm{unif}}(r)| & =&\sup
_{x}|H(F(x))-F(x)|\\
& =&\sup_{x}|G(x)-F(x)|.
\end{eqnarray*}
\end{appendix}

%s5 ###
\section*{Acknowledgments}
The authors thank Bradley Efron and Amir Najmi for useful comments
and discussion.
%the use of randomized $p$-values.}

%

%

%suskaldyti doi
% imsref loaded by akundreckaite, 2012-03-21 14:45:09

\printaddresses

\end{document}